\documentclass[showpacs,preprintnumbers,amsmath,amssymb,nofootinbib]{revtex4-2}
\usepackage{graphicx}
\usepackage{color, soul}
\usepackage{amsmath}
\usepackage{amsfonts}
\input epsf
\usepackage[pdftex,
            pdfauthor={A.A. Yurova, V. A. Yurov, A. V. Yurov},
            pdftitle={The cosmological arrow of time and the retarded potentials},
            pdfsubject={},
            pdfkeywords={Arrow of time; Klein-Gordon equation; Dirac equation; Schr\"odinger equation; retarded potentials; Hubble parameter},
            pdfproducer={Latex with hyperref},
            pdfcreator={pdflatex}]{hyperref}
\hypersetup{colorlinks=true, citecolor=blue, linkcolor=red}
\usepackage{enumerate}
\usepackage{scrextend}
\addtokomafont{labelinglabel}{\sffamily}

\newcommand{\beq}{\begin{equation}}
\newcommand{\beqn}{\begin{equation*}}
\newcommand{\enq}{\end{equation}}
\newcommand{\enqn}{\end{equation*}}

\newcommand{\R}{{\mathbb R}}

\newcommand{\Hh}{\mathcal{\hat H}}

\begin{document}
\title{The cosmological arrow of time and the retarded potentials}
\author{A. A. Yurova}%
\email{AIUrova@kantiana.ru}
\author{A.V. Yurov}%
\email{AIUrov@kantiana.ru}
\author{V.A. Yurov}%
\email{vayt37@gmail.com}
\affiliation{I. Kant Baltic Federal University, Department of Physics, Mathematics and IT, Al. Nevsky str. 14, Kaliningrad
236041, Russia}

\begin{abstract}
We demonstrate that the cosmological arrow of time is the cause for the arrow of time associated with the retarded radiation. This implies that the proposed mathematical model serves to confirm the hypothesis of Gold and Wheeler that the stars radiate light instead of consuming it only because the universe is expanding -- just like the darkness of the night sky is a side-effect of the global cosmological expansion.
\end{abstract}
% Keywords
%\keyword{Novikov-Veselov equation, Lax pair, Moutard symmetry, Cauchy problem}  % List three to ten pertinent keywords specific to the article, yet reasonably common within the subject discipline.

\maketitle
%%%%%%%%%%%%%%%%%%%%%%%%%%%%%%%%%%%%%%%%%%

%%%%%%%%%%%%%%%%%%%%%%%%%%%%%%%%%%%%%%%%%%

\section{Introduction}\label{Sec:intro}

It would not be an overstatement to claim that the nature of the arrow of time belongs to a small set of the most ancient and most pondered questions in physics and philosophy. To even attempt listing all the works and articles dedicated to this problem would be an exercise in futility -- so much has been written on the subject! The problem gets further obfuscated by the lack of common ground on what the term ``arrow of time'' actually means. The physicists usually define it as a physical distinction between what ``was'' and what ``will be'', an indicator that serves to distinguish and provide the formal definitions for the future and the past. This is as broad a definition as can be, and as such it suffers from the common problem plaguing all broad definitions: it allows for a number of {\em different} arrows of time! For example, Roger Penrose in \cite{RP-Molodetz} lists as many as {\em six} distinct time arrows, associated with: (i) a decay of $K^0$-meson, (ii) quantum measurements, (iii) growth of entropy, (iv) retardation of radiation, (v) our perception of time (a so-called psychological arrow of time), (vi) the expansion of the universe, and (vii) the disparity between the amounts of black and white holes. Alternatively, Steven Hawking in his bestselling book \cite{SH-Molodetz} restricts his attention to just three candidates: the time arrows (iii), (v) and (vi). So, even the classic disagree on the question of which of the arrows are more fundamental than the rest.

In this article we will take a new look on two of those time arrows, namely (iv) and (vi). We are going to show that these two share a very curious relationship. In particular, we will demonstrate that in the expanding universe there exists a global boundary condition which naturally favours retarded radiation to the advanced one. This appears to be a very intriguing and unexpected result, so much so that the authors themselves were quite taken aback by it. After all, we have always used to think about the cosmological arrow of time as a harmless curiosity and not a real physical principle. Perhaps we were too eager to condemn!

Interestingly, a possibility of a relationship between the arrows of time (iv) and (vi) has been discussed previously, in \cite{Hoyl-Narlikar-Molodetz}. The authors of that article have tried to establish a link between a local time arrow of electromagnetic field and a global cosmological time arrow within the framework of the Wheeler-Feynman absorber theory, which was very popular at the time \cite{Wheeler-Feynman-Molodetz-1}, \cite{Hogarth-Molodetz}. We are mentioning this to celebrate the historic justice and to give the due homage to our predecessors. We are going to use a more direct approach which does not require the absorber theory. But since we have already mentioned Richard Feynman, we would like to invoke a little fragment of his Nobel lecture, in which he recalls a conversation with German professor Herbert Jehle. They had a discussion about the possible applications of an action integral in quantum mechanics, and Jehle said: ``You Americans are always trying to find out how something can be used. That's a good way to discover things!'' \cite{Nobel}. Let us get in the full spirit of this quote by asking the following question: is it possible to put the derivation of the Pauli equation from the Dirac equation to some sort of use?

Here is the gist. The Pauli equation without the electromagnetic field is simply the Schr\"odinger equation for a free particle. But it is known that in order to properly add the interaction with electromagnetic field we have to use not the Schr\"odinger, but Dirac equation -- otherwise we would have no way to correctly calculate the magnetic moment of the electron. This has been known for a long time and it demonstrates that at least in one aspect -- the spin and the interactions with electromagnetic field -- the Dirac equation appears to be more fundamental than the Schr\"odinger equation+ for a free particle.

But what should happen if we decide to add the gravity field into the mix? We can do the same thing, only this time instead of grappling with the Dirac equation (as it was for the electromagnetic field) we shall dive even further, down to the Klein-Gordon equation (KGE). To be more precise, we will require all three equations: KGE, Dirac and Pauli (i.e. the Schr\"odinger equation for a free particle) equations, ordered as a hierarchy. The basic equation in this hierarchy will be KGE, from whom we can derive the Dirac equation (by ``taking a square root''), which in non-relativistic approach reduces to the Pauli equation -- or the Schr\"odinger equation for a free particle if there is no electromagnetic field. The reason we require this hierarchy and not just Schr\"odinger equation is simple -- the other two equations are needed for adding the {\em interactions with different fields}. To add an interaction with an electromagnetic field and producing a correct magnetic moment one has to start with the Dirac equation. Similarly, in order to properly add an interaction with a gravity we've got to start with KGE. This is due to the following observation. KGE is essentially a relativistic relationship between the momentum and the energy in the Minkowski space. By adding gravity we switch to pseudol-Riemann space which in terms of KGE simply means replacing the ordinary derivatives with the covariant ones. Once accomplished, we can then follow the standard algorithm for deriving the Pauli equation, i.e. take a square root and use the non-relativistic approach (by separating the large and small components of bispinor). As we shall see, by following this route already in the first approximation we will come to a very interesting observation: the gravity leaves a distinct mark in our calculations, selectively favouring the retarded solution, thus producing one of the most famous arrows of time -- the arrow of radiation.

Now, one can wonder: why such famous equations as KGE, Dirac and Schr\"odinger, themselves the subjects of countless first-rate works ~\footnote{Even the essential bibliography on the subject boggles the mind!}, have not been used in the way we propose in this article? All of these equations were extensively studied before, including the cases with variable gravity fields, but so far no one has made any claims about the time arrow allegedly hidden within. Why then do we dare to propose such an audacious proposition? The answer lies in cosmology. According to our contemporary models the universe we reside in must contain a fundamental scalar field. This field can be easily utilized for defining a time variable. Indeed, consider a cosmological hypersurface of a constant time, defined as a three-dimensional region of the universe with fields of matter identical in every point. These fields are commonly modeled by a single scalar field, which, once introduced, can play a role of a time variable. Take for example an inflaton field. A boundary exists which separates a metastable inflationary vacuum from the region where the vacuum state has decayed leading to a secondary reheating and a birth of a new pocket universe. It is this (infinite) boundary that defines a zero o'clock on every watch~\footnote{Or, shall we say, every pocket-watch?} of this pocket universe. Another important example involving the cosmological scalar field is the anthropic solution to the problem of a small cosmological constant. It is based on the idea, developed in \cite{VilGar-Molodetz} , which assumes that the density of the dark energy $\rho_{_D}$ is a random variable, and as such has to be written as a sum of two terms:
\begin{equation}
\rho_{_D}=\rho_{_{\Lambda}}+\rho_{\phi},
\label{VG}
\end{equation}
with $\rho_{_{\Lambda}} \in \R$ being the vacuum energy's density, and $\rho_{\phi}$ -- a variable density of a dynamical dark energy component, which in a simplest case act as certain fundamental scalar field $\phi$. Using (\ref{VG}) Vilenkin and Garriga have proposed an elegant solution to the problem of small observable cosmological constant which not only fits the observations (see \cite{Carter}--\cite{Weinberg}) but also naturally solves an adjacent problem of cosmic coincidences. For details we'll refer the interested reader to their article~\footnote{Interestingly, the very first paper which has suggested the anthropic solution of the cosmological constant due to the dark energy dates back to 1986 \cite{Linde-1}!}, but we will point out one important fact: the approach of Vilenkin and Garriga implies the existence of at least one global scalar field, which depends only on time (or, to be more precise, has negligibly small spatial gradients). An it is this field that we will implicitly use in eq. (\ref{Tmn}) in Sec. \ref{sec:Schrodinger}, which incidentally is the one equation that we will ``take a square root'' of.

The article is constructed in the following manner: the next three Sections will be dedicated to the arrows of time (iii), (iv) and (vi). In particular, Sec. \ref{sec:thermodynamics} is solely dedicated to the best known and widely accepted time arrows: the thermodynamic arrow of time, and also to its limitations in the cosmological framework. According to the modern theories of inflation, the post-inflationary universe automatically starts from a low-entropy state (which is good),  as long as we postulate an extremely low-entropy state of the universe {\em prior} to the beginning of the inflation (which is not so good). Thus, we essentially have to trade in the time arrow (iii) in the observable universe for the very unlikely and unexplained origins of the very early universe. And while this might still be explained and reconciled by a future quantum theory of gravity, for now the problem remains -- hence, its inclusion in Sec. \ref{sec:thermodynamics}, followed by the discussion of the time arrows (iv) and (vi) in Secs. \ref{sec:radiation} and \ref{sec:cosmology}. In Sec. \ref{sec:Schrodinger} we begin the calculations, deriving the Green's function and demonstrating its retardational properties. The results we discuss in \ref{sec:discussion}. In particular, we point out a steep price we had to pay for that result: the new modified Hamiltonians ends up being non-hermitian! This sounds quite radical and one might even wonder whether it actually discredits our entire approach. However, there is a physical meaning behind this complication: once the gravity and the cosmological horizon are included into the picture, the universe can no longer be considered a closed system; instead, one has to work with it as with any other open quantum system, which in general require complex Hamiltonians.

Before we move on, let us specify the assumptions that will be used throughout the article. There are three:
\begin{enumerate}
\item We restrict ourselves to the homogeneous and isotropic (Friedman) geometries;

\item We stick to the synchronous reference frame (the cosmic time) and the preferred coordinate system in which the cosmic background radiation is homogeneous and isotropic;

\item The Hubble parameter $H$ is assumed to be constant.
\end{enumerate}
We would like to point out that the third assumption does not actually imply us living in a de Sitter universe. Recall that the characteristic times of change for all cosmological parameters (such as scale factor $a$, Hubble parameter $H$, density $\rho$ and pressure $p$) are on the order of magnitudes of billions of years, whereas we are discussing the Schr\"odinger equation and the physics it describes on much smaller time scales. For that reason we can safely assume $H$ as a constant.

\section{Notes Found in the Bathtub and the Thermodynamic Arrow of Time} \label{sec:thermodynamics}

One of the most considered and reliable arrow of time is the thermodynamical arrow (iii), which {\em defines the future} as the direction of time along which the entropy grows. The thermodynamical arrow (iii) is stochastic in nature, but the probabilities involved are so large for the growth of the entropy and so small for its diminution, that the growing of entropy becomes practically predetermined. For illustration, consider an example from \cite{Susskind-Molodetz}. Suppose we stand in a bathroom that is completely and homogeneously covered by water vapors. What would be a probability that we shall see a small opening -- say, 1\% of the total volume of the vapors, -- spontaneously opening up in the steam? We can immediately see that it should be unlikely, since it would mean a decrease in entropy. Consider, for simplicity, that the steam is an ideal gas and that the process of opening up is isotermic. Then the entropy will vanish (in terms of Boltzmann units $k_{_B}$) by:
\begin{equation}
\Delta S=\frac{N}{99},
\label{entr}
\end{equation}
where $N$ is the amount of molecules of the water vapour. If this number is on level of magnitude of the Avogadro constant $N_{_A}$, we'll have the following probability:
\begin{equation}
P\sim \exp\left(-10^{22}\right).
\label{entr1}
\end{equation}
Looking at this tiny probability assures us that even we spend {\em millions of billions} of years in that bathroom we would not observe anything like this.

Boltzmann's arrow of time produced by the growth of entropy is a beautiful way to explain the past and the future. However, it does have one problem: in order to work it requires the initial entropy of the universe to be very low. For example, if we happen to be in the aforementioned bathroom and to our astonishment observe the small 1\% opening in the steam, we can be certain that in the future this opening will vanish. In fact, the probability of that with $1-\exp\left(-10^{22}\right) \approx 1$. We'll therefore have a clear distinction between the past state and the future state -- a proper arrow of time. Except we now have to answer the question of how did we get that low-entropy state (a homogenous steam with a small opening) in the first place. For isolated systems there exist two alternatives: (i) the bathroom was originally entirely covered by the steam, but a small fluctuation occurred, temporarily producing a small opening, only to be filled by the vapours again; (ii) in the more distant past the vacant space was even larger (say, $2\%$ of total volume), but at the moment of observation it was half-way done filling up by the water molecules. Normally we would choose (ii) with confidence, since usually the bathroom is free of steam before we open the faucets, and get progressively more opaque as the hot vapour began to flood the room. But this is true only because the bathroom with the vapours is an open system. If it was not, if the bathroom was an isolated system, then the alternative (i) becomes more preferable! Indeed, in this case the probability of alternative (ii) happening is equal to
\begin{equation}
P'\sim \exp\left(-2\times 10^{22}\right),
\label{entr2}
\end{equation}
so
\begin{equation}
\frac{P'}{P}= \exp\left(-10^{22}\right).
\label{entr3}
\end{equation}

And this is precisely the reason why the problem of low-entropy initial conditions for isolated systems is so difficult. In particular, the only way to explain the low observable entropy and highly structured content of our universe is to admit that it had had even smaller entropy at the very early stages of its life. In order to explain the seemingly well-ordered state of the current universe, one has to postulate an even {\em greater} order in the early universe! This looked like a blatant attempt at sweeping the problem under the carpet and for a long time such a state of affairs in cosmology was considered nothing short of scandalous. Only the advent of theory of cosmological inflation has cleared the picture. The inflation occurs very rapidly and it concludes by converting the energy of the inflaton field into the reheating and the birth of about $10^{80}$ elementary particles. The newly formed particle gas fills up the almost entirely flat (as a consequence of inflation) universe in a homogeneous and isotropic manner, producing what the classical thermodynamics would dub a high-entropy state. However, we have to account the gravitational degrees of freedom, and their presence completely changes the dynamics. The gas produced at the end of inflation is {\em almost} homogeneous, but it still has small lamps, tiny fluctuations in density -- the echoes of quantum fluctuations, blown up by the inflation. And from the point of view of gravitation this is a very low-entropy state, because it then proceeds to voraciously clumping those lumps together, eventually leading to formation of the galaxies, stars, planets and all the magnificent structure we can now observe. All of that increases the total entropy, just as it should according to the Second Law of Thermodynamics. So, we got a nice package in which cosmological inflation creates not only the universe we know but the thermodynamical arrow of time as well. However, this shining picture still has one small blemish: the problem of the {\em causes} of inflation. The inflation will only commence when there is a small but significantly isotropic and homogeneous volume of space filled with a metastable vacuum. Unfortunately, such a state is very low-entropy, and what is worse, we still have no substantiative evidences for such states being typical at small (say, Planckian) scales. In fact, it seems to be the opposite: at Plank scales we expect a very irregular space-time behaviour (quantum foam), which runs counter to the requirements for the inflation. Thus, even though the theory of inflation is pretty much universally accepted by the cosmologists, for now the initial conditions for the beginning of inflation seemingly belong to the set of ``unlikely'' and ``lucky'' occurrences. This is very regrettable, since this casts a long shadow on the theory of inflation being capable of single-handedly explaining the nature of thermodynamical arrow of time. Most physicists (including the authors of this article) hope that this mystery will be solved by an upcoming theory of quantum gravity. But for now we have to cope with a distinct possibility that the explanation for the arrow of time lies somewhere else.
%%%%%%%%%%%%%%%%%%%%%%%%%%%%%%%%%%%%%%%%%%%%%%%%%%%%%%%%%%%%%

\section{Retardation of radiation and Rock'n'Pond} \label{sec:radiation}

This was a particular point a view of Zeldovich and Novikov (ZN), who in 1975 have laid a bold claim that the distinction between the future and the past in terms of the growing entropy is fundamentally flawed \cite{ZelNov-Molodetz}. Their point of view was that the arrow of time have to exist in all physical systems, including, for example, the isolated pairs of particles for whom the entropy can not be properly defined. Even if somewhere far from a given pair there is some complex multi-particle system (such as the aforementioned bathroom with equimentioned steam) whose entropy can be defined, the local nature of physical laws would prevent it from exerting any kind of influence upon the happily isolated pair. One might argue, that the local laws of physics are time-reversible and thus cannot be used as litmus test for the direction of time~\footnote{For example, the Newtonian equations that govern the orbital movement of the planets in a Solar system are symmmetric w.r.t. the change of $t \to -t$. If somewhere in the universe there is a star system where the time is reversed, there would be no way of knowing it just by observing the rotation of the planets!}, but ZN respond that the real reason for the different between the future lies not the equations, but in the {\em initial conditions} that we impose upon those time-reversible equations. In particular, the conditions that produce either retarded or advanced radiation, while physically feasible, are utterly asymmetric with respect to the future and the past. In fact, ZN have argued that it is the condition that allows us to unequivocally distinguish the past from the future! Their idea was that the waves radiated {\em from} the emitter (retarded radiation) automatically correlate with its movement, whereas the {\em converging} waves (advanced radiation) require some very fine tuning of the frequencies and phases to completely give away their energy to the receiver. In other words, the retarded radiation comes chip (any accelerated charged particle can do it!), but emulating a time-inverted radiation is costly.

It is important to emphasize a subtle yet crucial difference between this approach and the claim that the observed dominance of the retarded radiation over the advanced one is just because the latter require very special, {\em low-entropy} boundary conditions. For example, when we throw a rock into a pond we casually observe the water waves travelling from the point of impact and beating upon the boundaries of the pond; the time-inverted process of the waves emerging from the pond boundaries, converging at the center of the pond and ejecting the rock in the air, while physically possible, requires extremely well-chosen boundary conditions, characterized by an astonishingly low entropy of the system ``water + boundaries of the pond + rock at the bottom''. This example might make us think that the arrow of time, associated with the retarded radiation must also be a by-product of the thermodynamical arrow of time (see Sec. \ref{sec:thermodynamics}). But ZN disagree, pointing out that even in the simple system of two interacting charged particles we shall get a retarded and not an advanced radiation. Since you cannot define an entropy of such a system, there can be no thermodynamic arrow of time. Thus, they conclude, it is the conditions for the radiation that defines the arrow of time, and it is more fundamental then the thermodynamic arrow of time. The entropy simply grows in the direction delineated by the condition of retarded radiation.

This reasoning sounds rather convincing, but like a proverbial pond it hides a rock at the bottom. When we consider an isolated system of two charged particles we are implicitly assuming them to be surrounded by an infinite empty space. We have to, because if they are surrounded by some other matter, no matter how distant, then the problem becomes similar to aforementioned rock thrown into a pond (or a lake) with very distant boundaries. And we have already discussed why this problem explicitly invokes the thermodynamic arrow of time. Thus, it looks like the argument of ZN is only true in the essentially empty universe, where the sole entertainment consists of the aforementioned two particles. Or is it?.. Do we actually know that in the empty universe there can be no converging waves reaching from the infinity?

On the first glance this sounds utterly ridiculous. But so is a particle (or two) as a unique occupant of the otherwise empty universe. Such a system is ripe for very unusual results. For instance, consider the example coined by Hoyle and Narlikar \cite{Hoyl-Narlikar-Molodetz}: a single particle with inert mass $m$ in an empty universe. Since there are no external forces, by the second Newton's law we have a null product of particle's acceleration and mass. What is in the solution? Hoyle and Narlika claims that the correct solution might actually be $m=0$, since there is no way to properly define the acceleration in an empty universe (no other particles = no points of reference). But then we got a truly Machian conclusion: the mass of a particle in a given universe is determined by all its material content. No content means no mass! An exceedingly poetic result, which is also utterly non-relativistic.

Thus, we reach an impasse. While the idea of ZN sounds solid, it is buildt using a model of very dubious physical relevance. Can we go so far as to claim that it is built on a shifting sand? Well, we argue that the answer is ``no''. But it will require some proof, which we will soon provide. Before that, however, there is one more arrow of time for us to discuss.

\section{The Cosmological Arrow of Time: When the Starts Steal the Light} \label{sec:cosmology}

We consider the cosmological arrow of time the last as it is the youngest and seemingly the ripest for criticism. Indeed, what do we mean by saying that the future is defined as the state of the universe with more distant galaxies? And does it imply that in the contracting universe the time goes ``backwards'', the children graduate from the college into the high school, argon atoms $^{40}$Ar spontaneously decays to the potassium atoms $^{40}$P, the black holes become white, and the stars suck in the light instead of emitting it? Sound like a hallucinogenic-induced nonsense, and yet it is what should in fact take place if the arrow of time is indeed produced by the expansion of the universe.

Here is the model originally proposed by Thomas Gold  \cite{Gold-Molodetz}. According to him, some time in the future the expansion of the universe might stop, and a new phase commence, during which the universe will contract and the entropy -- decrease. Gold then set an ambitious goal of finding a corresponding cosmological model that is completely time symmetric. Such a symmetry implies two entirely symmetric low-entropy boundary conditions -- one for the future, and one for the past. While we might balk at the thought of a future boundary condition, it is not logically inconceivable; after all, we are already used to the idea that the early universe requires a special low entropy boundary condition in the past, so why not the same thing in the future? Granted, as denizens of the expanding universe it is very difficult for us to imagine what the life in the contracting universe should look like (the time symmetry of the model ensures us that, yes, there has to be a life in such a universe). The pictures that comes to mind are staggeringly surreal: think of a steam being sucked into the water of a bathtub, or the black dots of stars on the luminescent skies, or the old people getting younger with every year. Imagine the eyes of the observer emitting the light which is sucked up by the burning candle, wax stains climbing up its sides, while the observer recalls the future and ponders the past... Such a place would put to shame even the Hogwarts and Wonderland combined. And yet no denizen of the collapsing universe would see them! According to Gold, the arrow of time in such a universe must be reversed, which means that these people would see the universe expanding, and entropy growing. In fact, it is us they would pity as the weird ones who live in the collapsing universe!

Naturally, most cosmologists thought of Gold's model as too exotic for their tastes, and claimed there really no real physical reasons to impose any special boundary conditions in the future (aside from the obvious aesthetic beauty stemming from the symmetry of the model). Stephen Hawking once attempted to formulate one such one reason in \cite{Hawking-Molodetz}, but he withdrew his claim shortly afterwards. Furthermore, most contemporary cosmologists stick to a view that it is unlikely for our universe to ever stop expanding, and even if it should \cite{VilGar-Molodetz}, the subsequent collapse would be totally unlike the time-inverted phase of expansion. All of this seems to make the Gold's model and, by extension, the cosmological arrow of time dubious at best and outdated at worst. Zeldovich and Novikov had a field's day with it in \cite{ZelNov-Molodetz} by mercilessly ridiculing the claim that the arrow of time in a closed universe must invert once the universe starts contracting. They argued that the cosmological switch from expansion of the universe to contraction is completely analogous to the behaviour of a vertically propelled rock (or a space rocket with the velocity smaller than the orbital speed) when the force of gravity stops its upward movement and compels it to descend. The rock experiences no change in the direction of time, and neither should the universe. Thus, concluded ZN, it makes no sense whatsoever to use the expansion of the universe as a fundamental arrow of time. They've even apologized to the readers afterwards for spending so much time on ``explaining the elementary'', which can only be excused by so many articles ``persisting in perpetuating the flawed arguments'' \cite{ZelNov-Molodetz}.

With all due respect to ZN as two indubitable classics, we have to point out that the argumentation of theirs is equally ``flawed'', and the ``flaw'' stems from the false analogy. The movement of the rock and the evolution of the universe are two very different physical phenomena and one cannot expect to simply use one to explain another. It is very easy to see, once we recall that the rock (or, more generally, a rocket) moves along the geodesics according to the laws of relativity -- in particular, it can never move with the velocities exceeding the speed of light. On the other hand, the expansion of the universe is produced by a literal {\em stretching} of the space-time, and is not bound by the speed of light~\footnote{A distance between two significantly distant objects in a rapidly expanding universe might grow faster than the speed of light. According to the observational data, ours is one such universe, cf. \cite{Perlmutter},\cite{Riess}}. Hence, despite some surface-level similarities, the fundamental difference in the physical nature of these two cases makes any kind of comparison by analogy dubious to say the least. We apologize to our readers for wasting so much time on something so elementary, but in our opinion it was necessary to point out an obvious flaw in ZN's argumentation.

Still, one might retort that a flaw or no flaw, we still don't know what happens with the direction of time in the contracting universe. Can we provide any sort of tangible proof that the the cosmological dynamics has any effect et all upon the arrow of time? As a matter of fact, we do! In fact, we are going to show that the electromagnetic time arrow (iv) is in indeed predetermined by existence of cosmological arrow (vi), so that in the collapsing universe the stars are indeed sucking the light out of outer space! But before we begin with our investigation we have to first answer one general question: how can the expansion of the universe have any effect upon the retarded radiation such as the one emitted from our smartphones when we turn the flashlight on? One is a global process that manifests itself on the level of galaxies and above, while the other is a product of a local physical laws. How can one affect another? The answer lies in a {\em global boundary condition}.

To understand it, consider the field equations derived by the variational principle. In order to to get to them, we have to variate the action which is followed by integration by part \cite{Landau1}, \cite{Landau2}. This produces the Lagrange-Euler equation and an additional surface term. In order to dispense with the latter we impose the following global condition: both the fields and their derivatives on the infinitely removed hypersurface must be equal to zero. Hence, even though in the end we get local equations, they are all predicated upon the existence of one global boundary condition. And it is this almost obvious observation that is going to loom large in the next Section.

\section{The Schr\"odinger equation in expanded universe} \label{sec:Schrodinger}

We shall start our derivation with the Klein-Gordon equation for scalar field with no gravity:
\begin{equation}
 {\ddot\phi}-c^2{\Delta}\phi+\left(\frac{mc^2}{\hbar}\right)^2 \phi=0,\qquad
\Delta\phi=\frac{\partial^2\phi}{\partial x^2}+\frac{\partial^2\phi}{\partial y^2}+\frac{\partial^2\phi}{\partial z^2}.
\label{KG}
\end{equation}
Physically speaking, (\ref{KG}) is a relativistic relationship between the momentum and energy in the Minkowski space-time. As such, it shall hold for any free (non-interacting) relativistic fields. The Dirac equation can be derived if we formally ``take a square root'' of (\ref{KG}).

Next, let us make an upgrade and go from the Minkowski space to the more interesting geometry of Friedmann universe with scale factor $a=a(t)$. The analogue of (\ref{KG}) for this case can be obtained using the fact that the covariant derivative of the stress-energy tensor is equal to zero (as a consequence of contracted Bianchi identities):

\beq\label{Tmn}
T^{\mu\nu}_{;\mu}=0\to {\ddot\phi}+3\frac{\dot a}{a}{\dot\phi}-\frac{c^2}{a^2}{\Delta}\phi+M^2\phi=0,
\enq
or, speaking more precisely, we get a product of (\ref{Tmn}) and $\dot\phi$, but it is not important here \footnote{This becomes important when one describes the phenomena of the phantom zone crossing}. For simplicity, let us rescale the spatial variables as such:
$$
\frac{1}{a}{{\partial}_i}\to\partial_i,\qquad i=1,2,3.
$$

%On our next step, let us ``take a square root'' of (\ref{Tmn}).
On our next step, we have to reduce the order of the second order differential equation (\ref{Tmn}) by build up a corresponding $4
\times 4$ matrix differential equation of first order. This can be done properly in a number of ways. For instance, one can adopt the factorizational approach described in \cite{Bogoliubov}, which is commonly used for turning \eqref{KG} into a Dirac equation. However, in our case we decided to follow a simpler route, delineated in a wonderful monograph \cite{Bjorken}. Here is how the method works for \eqref{KG}. First, we write down a linear equation containing a first order derivative w.r.t. variable $t$ and define the Hamiltonian $\Hh$ which contains the spatial derivatives of no higher than first order:
\beq \label{Bjorken}
i \hbar \frac{\partial}{\partial t} \Psi=\Hh \Psi.
\enq
The coefficients in $\Hh$ must be $4\times 4$ matrices. Taking a square of \eqref{Bjorken} \cite{Bjorken} produces a new equation:
\beq \label{Bjorken-squared}
- \hbar^2 \frac{\partial^2}{\partial^2 t} \Psi=\Hh^2 \Psi.
\enq
Now all that remains would be to require every component of \eqref{Bjorken-squared} serving as a solution of KGE \eqref{KG}. This both defines the unknown coefficients and yields the Dirac equation we required.

What if instead of equation \eqref{KG} we start out with \eqref{Tmn}? The crystal clear logic of \cite{Bjorken} is still fully applicable here. Repeating all the steps this time yields a new matrix equation, which is very similar to a Dirac equation expect for one term depending upon the Hubble parameter $H = \dot a / a$:
\begin{equation}
i\hbar \left(\frac{\partial}{\partial t}+\frac{3}{2}H\right)\Psi=-i\hbar c\sum_{k=1}^3\alpha_k\partial_k\Psi+mc^2\beta\Psi,
\label{Dir}
\end{equation}
where the constant matrices satisfy the usual anticommutative algebra
$$
\{\alpha_i,\alpha_k\}=2\delta_{ik},\qquad
\{\alpha_i,\beta\}=0,\qquad \alpha_i^2
=\beta^2=E.
$$
and can thus be written in terms of the Pauli matrices:
$$
\alpha_k=\left(\begin{array}{cc}
0 & \sigma_k\\
\sigma_k & 0
\end{array}\right),
\qquad \beta=\left(\begin{array}{cc}
E & 0\\
0 & -E
\end{array}\right),
$$
Note also that the mass in (\ref{Tmn}) also depends upon $H$ as follows:
$$
M^2=\left(\frac{3H}{2}\right)^2+\left(\frac{mc^2}{\hbar}\right)^2.
$$

Before we move on, we would like to stress that, just like the original Dirac equation, the modified equation \eqref{Dir} is {\em invariant w.r.t the reversal of time}. In order to see this, first recall that the time reversal automatically alters the sign of the Hubble parameter $H=\frac{d}{dt}(\ln a)$ (for those who go backwards in time, the universe appears to be contracting). So, let us define:
\beq \label{t_to_t'}
t' = t, \qquad \partial_k'=\partial_k, \qquad H'=-H.
\enq
A time-inverted solution of \eqref{Dir} can be obtained from $\Psi(t)$\footnote{For the sake of brevity we omit the spatial variables in the argument of $\Psi(t,\vec r)$} by the operation of complex conjugation followed by additional multiplication by a certain $4\times 4$ matrix $T$ \cite{Bjorken}:
\beq \label{Psi-T}
\Psi'(t')=T \Psi^*(t),
\enq
which produces the following time-inverted version of \eqref{Dir}
\beq \label{DirInverted}
i\hbar \left(\frac{\partial}{\partial t'}+\frac{3}{2}H'\right)\Psi'(t') = \left(-i\hbar c\sum_{k=1}^3 T \alpha^*_k T^{-1}\partial_{k}' + m c^2 T \beta^* T^{-1}\right) \Psi'(t'),
\enq
where the matrix $T$ can be shown to be equal to
\beqn
T=-i \alpha_2 \alpha_3 = - \left(
\begin{array}{cc}
\sigma_2 & 0\\
0 & \sigma_2
\end{array}\right),
\enqn
because $T$ shall commutate with $\alpha_2$ and $\beta$, but anti-commutate with $\alpha_1$ and $\alpha_3$. And if we revert from $\Psi'$ back to $\Psi^*$ by \eqref{Psi-T}, take a
complex conjugate of \eqref{DirInverted}, and return to the original $t$ and $H$ as in \eqref{t_to_t'}, we'll get nothing else but our equation \eqref{DirInverted}, as it ought to be.

Next, we should use the nonrelativistic approximation. In order to do that it is handy to rewrite (\ref{Dir}) as:
$$
i\hbar \frac{\partial}{\partial t}\Psi=-i\hbar c\sum_{k=1}^3\alpha_k\partial_k\Psi+\left(mc^2\beta-\frac{3iH}{2}\right)\Psi,
$$
after which we can express the bispinor in terms of two spinors via classic substitution
\begin{equation}
\Psi={\rm exp}\left(-\frac{imc^2}{\hbar}t\right)\left(\begin{array}{c}
\phi\\
\chi
\end{array}\right).
\label{bisp}
\end{equation}
Then
\begin{equation}
 i\hbar \frac{\partial}{\partial t}\left(\begin{array}{c}
\phi\\
\chi
\end{array}\right)=
-i\hbar({\vec\sigma}{\vec\nabla})\left(\begin{array}{c}
\chi\\
\phi
\end{array}\right)-
2mc^2\left(\begin{array}{c}
0\\
\chi
\end{array}\right)-
\frac{3i\hbar H}{2}\left(\begin{array}{c}
\phi\\
\chi
\end{array}\right),
\label{bi2}
\end{equation}
Then, separating the large and small bispinor components and after a few simple calculations we end up with our desired equation:
\begin{equation}
i\hbar\frac{\partial\psi}{\partial t}=-\frac{\hbar^2}{2m+i\lambda}\Delta\psi,\qquad \lambda=\frac{3\hbar H}{2c^2}.
\label{zap1}
\end{equation}
This is the sought after generalization of Sch\"odinger equation for a free particle which travels in the expanding ($H>0$) Friedman universe. And this equation has an interesting special point in the lower complex half-plane. As we shall soon see, it is this peculiarity that ensures that the Green's function for (\ref{zap1}) in the expanding universe will always be {\bf retarded}. But before we get there, we should emphasize one important aspect of \eqref{zap1}: just like its predecessor \eqref{Dir}, this equation is {\em invariant w.r.t. the time inversion}! Indeed, similar to an ordinary Schr\"odinger equation, the time reversed version of \eqref{zap1} can be obtained by replacing $t \to -t$, $\psi \to \psi^*$ and, of course, $H \to -H$. The latter is crucial, because it guarantees that the modified Schr\"odinger equation \eqref{zap1} remains indifferent to a direction of time -- as long as we remember that, in the optics of those living backwards in time, an expanding universe is actually collapsing! In other words, by deriving the equation \eqref{zap1} from \eqref{Dir}, we have not accidentally smuggled in an arrow of time into the model that {\em had none}. This would have amounted to nothing short of a mathematical error! And yet -- the time arrow is now very close by, its contours ready to take shape once we add a final piece of the puzzle: the fact that the observed universe is {\em expanding}.

So, let's get on with it! In order to prove that there is no foul play and no aces up the sleeves of the authors, let us slowly walk through calculations along with the reader.

Let us start by recall that the Heaviside function $\theta$
\begin{equation}
 \theta(\tau>0)=1,\qquad \theta(\tau<0)=0,
\label{hav0}
\end{equation}
which satisfies the following relationship:
\begin{equation}
\displaystyle{
\int_{-\infty}^{\infty} \frac{d\omega}{\omega+i\epsilon}{\rm e}^{-i\omega\tau}=-2\pi i\theta(\tau){\rm e}^{-\epsilon\tau}.
}
\label{hav}
\end{equation}
where $\epsilon>0$. The relationship (\ref{hav}) is commonly used as an integral representation for (\ref{hav0}) in the limit $\epsilon\to 0$. (\ref{hav}) can be proven easily by closing the contour of integration in the lower complex half-plane, computing the resulting integral by the residue theorem at point $\omega=-i\epsilon$ \cite{Ahlfors}. Assuming that the lower contour stretches up to a negative infinity, for $\tau > 0$, the integral over the lower contour will vanish and we will get (\ref{hav}). When $\tau<0$ we shall close the contour in the upper complex plane; since there are no poles there, the integral will be equal to zero.

Now let us concentrate on the Green's function. By definition, it is the function $G$ that satisfies the equation:
\begin{equation}
\displaystyle{
\left(i\hbar\frac{\partial}{\partial t'}+\frac{\hbar^2}{2m+i\lambda}\,\Delta'\right)G(x'-x)=\delta^{(4)}\left(x'-x\right),}
\label{zap2}
\end{equation}
where
$$
\delta^{(4)}\left(x'-x\right)=\delta\left(t'-t\right)\delta^{(3)}\left({\bf r}'-{\bf r}\right),\qquad G(x'-x)=G(t'-t,{\bf r'}-{\bf r}).
$$
and the boldface is used whenever we meet with a three-dimensional vector. The Fourier transformation of $G$ is
\begin{equation}
\displaystyle{
G(x'-x)=\int \frac{d\omega}{2\pi}\frac{d^3p}{(2\pi\hbar)^3}{\rm e}^{-i\omega(t'-t)+i{\bf p}({\bf r'}-{\bf r})/\hbar}G_0({\bf p},\omega),}
\label{zap3}
\end{equation}
and if we substitute (\ref{zap3}) into (\ref{zap2}), take into account an explicit form of the Fourier transform of $\delta^{(4)}\left(x'-x\right)$, we end up with the following equation for $G_0({\bf p},\omega)$:
\begin{equation}
\displaystyle{
G_0({\bf p},\omega)=\frac{1}{\hbar\omega-\frac{2m{\bf p}^2}{4m^2+\lambda^2}+\frac{i\lambda {\bf p}^2}{4m^2+\lambda^2}}\equiv \frac{1}{\Omega+i\epsilon},}
\label{zap4}
\end{equation}
where we have introduced two auxiliary parameters $\Omega$ and $\epsilon$.  We can now substitute (\ref{zap4}) back into (\ref{zap3}) and integrate over $\omega$ using (\ref{hav}). This results in
\begin{equation}
\displaystyle{
G(x'-x)=-\frac{2\pi i}{(2\pi\hbar)^4}\theta\left(\frac{t'-t}{\hbar}\right)\int_{-\infty}^{\infty}d^3p\exp\left(\frac{{\bf p}^2(t'-t)}{\hbar(2im-\lambda)}+\frac{i{\bf p}({\bf r'}-{\bf r})}{\hbar}\right),}
\label{zap5}
\end{equation}
And we conclude our calculations by pointing out that, according to (\ref{zap5}), the Green's function is indeed {\em retarded}:
\begin{equation}
G(x'-x)=0,\qquad t'<t
\label{VOT}
\end{equation}
and it is indeed {\em caused by an expansion of the universe}: $H>0$, i.e. $\epsilon$ in (\ref{zap4}) belons to a lower complex half-plane. It follows automatically, that in the collapsing universe the propagator must be {\em advanced} -- just as Gold envisioned  using his hypothesis of ``symmetric universe''. But unlike his, our result is strictly necessitated by mathematics behind the derivation of the Schr\"odinger equation (\ref{zap1}).

For the sake of completeness let us finish the calculations by taking the remaining few integrals. For this end, let us choose the ${\bf p}_z$ axis along the ${\bf r'}-{\bf r}$, switch to spherical coordinates:
$$
\{p_x,\,p_y,\,p_z\}\to \{0\le p<\infty,\,0\le \theta\le\pi,\,0\le\varphi\le 2\pi\},
$$
and integrate w.r.t. $\theta$. Then choose new scalar variable $k={\bf p}({\bf r'}-{\bf r})/\hbar$. The expression would then convert to:
\begin{equation}
\displaystyle{
G(x'-x)=-\frac{ i}{(2\pi)^2\hbar |{\bf r'}-{\bf r}|^3}\theta\left(\frac{t'-t}{\hbar}\right)
\int_{-\infty}^{\infty}dk\, k \sin k\exp\left(-\frac{\hbar (t'-t)k^2}{|{\bf r'}-{\bf r}|^2(\lambda-2im)}\right).}
\label{zap6}
\end{equation}
By calculating the remaining integral we get the final result:
\begin{equation}
\displaystyle{
G(x'-x)=-\frac{i}{\hbar^{5/2}}\left(\frac{3i\hbar H+4mc^2}{8i\pi c^2 (t'-t)}\right)^{3/2}  \theta\left(\frac{t'-t}{\hbar}\right)
\exp\left[\left(\frac{im}{\hbar}-\frac{3H}{4c^2}\right)\frac{|{\bf r'}-{\bf r}|^2}{2(t'-t)}\right].}
\label{final}
\end{equation}
It is obvious that, as expected, in the special case $H=0$ this expression reduces to a standard retarded Green's function for a free non-relativistic particle.

Thus, we have demonstrated the existence of an unlikely yet strong relationship between the cosmological arrow of time in the expanded universe and the arrow of time produced by the retarded radiation. The observant reader might notice that we have derived the retarded propagator for just a single particle, but this can be easily remedied using the same approach. The result would still have to account for the same pole in the lower complex half-plane, which once again would favour the retarded propagator to the advanced one. And this, of course, produces the classic retarded radiation. According to our model, because the electromagnetic wave is formed of photons, and because the wave functions of these photons share the same retarded Green's function -- a propagator in the expanded universe, -- we get an electromagnetic arrow of time. In other words, we get an explanation for why the stars shine instead of consuming the nearby light, or why the eye of the reader sees this article instead of projecting the photons out of a retina back onto the monitor. And what is even more fascinating -- we get this free of charge, simply because of the expansion of the universe! And yes, it is indubitably ironic if all the night lanterns work just because the universe happened to expand; then again, it only gets dark at night {\em because} of that expansion, so we are in no position to complain \cite{Harrison}!

Before we conclude this Section, we should say a few words about the meaning of our result. In particular, it is interesting to ponder whether it applies solely to an electromagnetic arrow of time or, more broadly, to the time arrow of radiation in general? We would argue that the latter is true. If we think about it, the water waves in the pond discussed in Sec. \ref{sec:radiation} or, say, the longitudinal acoustic waves in the air are merely the macroscopic manifestations of movement of numerous numbers of atoms, fully controlled by the quantum mechanical laws and equations -- the very equations that ought to contain a tiny imaginary term just like in (\ref{zap1})\footnote{If the reader is of a mind that the quantum mechanics has no hand in these phenomena, we fully recommend an interesting article by Andreas Albrecht and Daniel Phillips \cite{Albrecht-Molodetz}, which vividly demonstrates the tremendous rate at which the quantum uncertainty becomes a dominant factor during the collisions of ordinary particles. For example, the behaviour of the Nitrogen molecules at standard air temperature and pressure is essentially non-classical at all times, whereas for the water molecules taken at body temperature the quantum uncertainty overcomes the classical randomness already after the first collision. Even for something as inconspicuous as a game of billiards (or a snooker), on average the eight collisions on a pool table is enough for claiming that the random movement of a billiard ball is now governed by an overgrown quantum uncertainty! Naturally, this applies with a vengeance to all the phenomena that involve huge numbers of colliding molecules, including both the water and the air waves.}. Thus, the evolution of each of these atoms must be governed by a retarded propagator. An apt analogy here would be the famous double-slit experiment. Imagine an opaque plate with two open slits. If we illuminate this plate with a coherent source of light (a laser), on the screen behind the plate we will observe a certain interference pattern. However, if we pierce the plate twice more, adding two more slits, a new pattern will form, and some previously illuminated places on a screen will no longer register any photons whatsoever. One might at first interpret it as a collective effect and argue that the new slits produce two new streams of photons that play havoc with the photons from the original two streams, somehow hampering their progress to certain places on the screen. But it cannot be! One might repeat the same experiment with a single photon, and it would still shun the same places on the screen. Hence, the quantum interference has nothing to do with a collective behaviour of photons, and everything -- with how a single photon behaves itself. And it is by analogy with this famous experiment  that we propose a following hypothesis: the time arrow we have derived must be pertinent for every single atom and it is through {\em them} that the effect of a time arrow is felt by the ``classical'' waves, be it a water wave, an air wave, or an electromagnetic wave. Or, to put it in other words, the time arrow of a wave is comprised of multiple time arrows of its constituent atoms.

\section{Discussion} \label{sec:discussion}

The modified Schr\"odinger equation (\ref{zap1}) has allowed us to come to an interesting conclusion about the interdependence of two different arrows of time. But it also invites a number of questions, such as: does it imply that the Hamiltonian in the expanded universe is no longer hermitian? And isn't it a tad too steep a price for our endeavours?

Let us consider it. We know that at $H=0$ (and thus, at $\lambda=0$) the simplest solution of (\ref{zap1}) is a spherical wave:
\begin{equation}
\psi(r,t)=\frac{{\rm e}^{-iEt/\hbar+ikr}}{r},\qquad k=\frac{\sqrt{2mE}}{\hbar}.
\label{sfer}
\end{equation}
A not very large non-zero $H$ will alter this definition but slightly, due to an extremely small coefficient $\hbar/c^2$. Letting $\lambda\ne 0$, and assuming that the energy is real-valued, we shall get
\begin{equation}
k=\frac{\sqrt{(2m+i\lambda)E}}{\hbar}.
\label{kk}
\end{equation}
Let us estimate the parameters involved. Using the currently observed value for the Hubble parameter $H\sim 2.2\times 10^{-18}$ $s^{-1}$ (see \cite{CKR} for an interesting discussion on the discrepancies in determining the values of $H$) we can see that $\lambda=2.42\times 10^{-65}$ g, which is by 37 orders of magnitude (!) lighter than the mass of the electron, $m_e=9.1\times 10^{-28}$ and by 31 orders of  magnitude lighter than neutrino, the lightest known particle, $m_{\nu} \approx 2.14 \times 10^{-34}$ g! Therefore, we can rewrite (\ref{kk}) with a very high accuracy as
\begin{equation}
k=\frac{\sqrt{2mE}}{\hbar}\left(1+\frac{i\lambda}{2m}\right).
\label{kk1}
\end{equation}
Hence, the solution (\ref{sfer}) will get one additional exponential multiple. Setting $E=mc^2$ and introducing a horizon radius
$$
R_{_H}=\frac{c}{H},
$$
allows us to write this multiple as
\begin{equation}
\exp\left(\pm \frac{3\sqrt{2}}{4}\frac{r}{R_{_H}}\right).
\label{vot}
\end{equation}

For instance, at the scale of a standard lab, $r\sim 100$ meters, this multiple in (\ref{sfer}) will be equal to $1\pm 7778\times 10^{-26}$. In other words, a presence of a tiny imaginary ``cosmologically induced'' term in (\ref{zap1}) is, for all intents and purposes, virtually imperceivable in the lab. Its only measurable contribution will be an emergence of a time arrow under the guise of retarded radiation!

However, a deviation from self-adjacency, even as tiny and imperceptible as this, still requires a physical explanation, otherwise it will be deemed unacceptable by the absolute majority of physicists (including the authors of this article!). Fortunately, the equation (\ref{zap1}) actually makes a lot of sense in the framework of a classical quantum mechanics. First, we have to understand that the non-stationary cosmology effectively invalidates the common view on the the universe as a closed system \cite{LandauStat}. Instead one has to consider it as a system embedded in a variable gravitational field -- note, that one cannot just add this field as a component of a system lest all the conservational laws turn into trivial identities, losing us the key tools of statistical approach. Hence, we are dealing with an {\em open system}, i.e. a system interacting with an environment. Such systems are well-known and well-studied. For example, they commonly arise in the quantum theory of dissipative processes (there the word ``environment'' is usually replaced by a term ``reservoir'') and in the description of the physical nature of quantum measurements. The methods used for probing the open systems are numerous and include: the Lindblad generalization of von Neumann equation \cite{Lind-Molodetz}; the quantum diffusion model of Caldeira and Leggett \cite{Kal-Molodetz}; the quantum theory of continuous measurements, based on the Feynman integrals \cite{Fey-Molodetz} and further developed by Mensky in \cite{Men-Molodetz}. The latter approach is most pertinent for us as it naturally leads to complex Hamiltonians of appropriate form. Let us take a short look at this approach (as usual, for more details we refer the reader to the aforementioned articles).

Consider a variable which describes a certain state of an open system. Let us assume that this variable gets continuous varied due to external influence upon the system, and that its resulting value is denoted by $\alpha$ . Then we can introduce a measure $w_{\alpha}[p,q]$ which depends on the trajectory in a phase space and is close to $1$ if the trajectory is close to $\alpha$ but is almost zero for all the other trajectories. The evolution of the system we can describe via the Feynman integral:
\begin{equation}
\displaystyle{
\int [dp][dq]w_{\alpha}[p,q] \exp\left[\frac{i}{\hbar}\int_0^t dt'\left(p{\dot q}-{\mathcal H}(p,q,t')\right)\right],}
\label{propagator}
\end{equation}
Next, let us specify what we mean by a measurement. Suppose we are observing the variable $A$, and the result of observation is denoted by $a(t)$. Then the bounded Feynman integral must be taken over a narrow ``corridor'' of possible trajectories, centered around the one trajectory which produces the $a(t)$. It is possible to show that the measure functional $w_{a}[p,q]$ shall be chosen in a form of a Gauss functional:
\begin{equation}
\displaystyle{
w_{a}[p,q]=\exp\left[\int_0^t dt'(-\kappa\left(A(p,q)-a(t')\right)^2)\right],}
\label{propagator1}
\end{equation}
where $\kappa$ is a coefficient which controls the resolution. This produces a modified Schr\"odinger equation with an effective complex Hamiltonian:
\begin{equation}
|{\dot\psi}\rangle=-\frac{i}{\hbar}{{\Hh}_{eff}}=\left(-\frac{i}{\hbar}{\Hh}-\kappa\left({\hat A}-a(t)\right)^2\right)|\psi\rangle.
\label{eff}
\end{equation}
which we have written using the notation introduced in \cite{Men-Molodetz}. It is now easy to check that (\ref{zap1}) is reducible to (\ref{eff}):
\begin{equation}
\begin{array}{l}
{{\Hh}_{eff}}={\Hh}-i\kappa \left({\hat A}-a(t)\right)^2=\\
\\
-\frac{\hbar^2}{2m+i\lambda}\Delta=-\frac{\hbar^2}{2M}\Delta-i\frac{\lambda}{4m^2+\lambda^2}\left(-i\hbar\nabla-0\right)^2.
\end{array}
\label{vot}
\end{equation}
if we redefine the mass as
$$
m\to M=m+\frac{\lambda^2}{4m}.
$$
Therefore, in the terms of open systems, the equation (\ref{zap1}) simply describes a continuous observation of a particle's momentum within a quantum ``corridor'' centered around the zero momentum. This is a fascinating picture which provides an interesting illustration to a possible physical explanation for a non-hermitian form of our Hamiltonian, as well as stimulates an interest in further investigation of its nature.

Before we move on to a final conclusion, a couple of remarks is in order. First of all, we must acknowledge that the relationship between the observed asymmetry of radiation and the retarded/advanced solutions of the wave equations might actually be very complex and non-trivial. This is the subject that we did not even dare to touch, once we took into account the wide range of different opinions among the authors who write on the topic, well illustrated by Hew Price in \cite{Price-Molodetz}. We do not deem ourselves ready for tackling this difficult subject, but harbour a hope that we will return to it in a later article. The second question we have not discussed so far is the problem of a time arrow in a universe that begins to collapse, especially in the instant when the Hubble parameter changes sign. This question becomes surprisingly relevant if we are correct in our assumptions about the existence of a global scalar field. According to J. Garriga and  A. Vilenkin  in \cite{VilGar-Molodetz}, the (positive) density of a scalar field decreases in time whereas the density of the vacuum energy remains negative. In a distant future this trend will lead to effective cosmological constant changing sign and becoming negative, ensuring the imminent end of expansion of the observable universe, and heralding its subsequent collapse. The qualitative estimates made in \cite{VilGar-Molodetz} states that it will happen after $10^{12}$ yr, although the time required might be cut short by the unusual singularities discovered by J. D. Barrow and  A. A. H. Graham in \cite{BarrowSuperMolodetz}, reducing it to something comparable with an age of contemporary universe \cite{Alla-Lera-Tema-Molodetz}. But if there indeed exists a direct link between the cosmological and radiational arrows of time, then the collapsing universe will be dominated by the advanced (not retarded!) radiation. What would it mean for an observer in such a universe? This is a good question which we are studying right now and are expecting to discuss in a subsequent paper. Finally, we have left open an interesting question about the relationship between the cosmologic and thermodynamic arrows of time. The brightest example of it are the stars~\footnote{We are grateful to one of our referees who brought this example to our attention}. From the common point of view, the stars are the places where the thermodynamic gradient is particularly steep, so the temporal asymmetry of radiation can be considered a consequences of the thermodynamic asymmetry. However, we are leaning to a different point of view, which, despite our critical remarks in Sec. \ref{sec:radiation} we (at least) partially share with ZN. If our hypothesis is correct, then the three different arrows of time form a following hierarchy:
\begin{enumerate}
\item The cosmological arrow of time. It is the most fundamental of the three as it ensures the physical asymmetry between the future and the past.

\item The cosmological arrow of time produces the time arrow of radiation as described in Sec. \ref{sec:Schrodinger}.

\item Once the physical asymmetry between the past and the future is established, we can begin to formulate the Second law of thermodynamics and postulate that the entropy always grows in the direction of the future.
\end{enumerate}

According to ZN, the growth of entropy is a handy tool for probing the direction of time, as long as we remember that it is just a by-product of a more fundamental time arrow. And while we agree with ZN about this, we disagree with their opinion as to what is more fundamental. They thought that it was 2 -- the time arrow of a radiation. We argue that it is in turn an offspring of a more global time arrow 1, which takes root in nothing else but an expansion of the universe!

In conclusion, we would like to once again emphasize that we are as astonished as anyone by these results. We have always been on the side of Zeldovich and Novikov with regards to the question of a validity of cosmological arrow of time. Up until now, we simply did not believe in it as something even vaguely fundamental -- unlike the well-established thermodynamic arrow of time. However, once we have started working on this article, we began to encounter more and more additional arguments for a cosmological nature of the direction of time. For example, consider a pure de Sitter universe, devoid of any matter. The astronomical data forces us to believe that this foreboding model is a dark and extremely distant future of the universe we live in -- so distant, in fact, that the cosmic period during which the universe remains filled with the sparkling matter of stars, galaxies -- or even plain baryon matter, -- is merely a measure zero blip in the overall timeline.~\footnote{If the vacuum with a positive energy density is metastable, the universe will have an expiration date and will not exist forever. However, an estimated duration of time prior to decompactification is still so large, that it not just dwarfs the aforementioned time period, it ``quarkifies'' it!} In a rather short time (cosmologically speaking) a cosmic event horizon will contain nothing at all except for Hawking-Gibbons radiation (even the cosmic background radiation will be a distant memory in just a trillion years, too faint to be detected). The universe will inevitably slip into a state of a maximal entropy. However, the very presence of the Hawking radiation implies there will exist slight temperature fluctuations. Such fluctuations will temporarily decrease the overall entropy. They will dissipate, of course, rising the entropy back to the max. The problem is -- both the {\em prior} and the {\em succeeding} states will have more entropy than the universe had at the moment of fluctuation. This implies that a de Sitter universe has no discernible thermodynamic arrow of time. And yet there would exist at least one other time arrow, tirelessly measuring eon after eon -- and that, of course, would be the cosmological arrow of time! So, does it mean that it actually {\em is} more fundamental? We do not know. Hopefully, in the future we will be able to answer this question. After all, in this article we did not presume to tell the future -- merely to tell it from the past.

\section*{Acknowledgments}
The article was supported by the Ministry of Science and Higher Education of the Russian Federation (agreement no. 075-02-2021-1748).


\begin{thebibliography}{999}
% Reference 1
\bibitem{RP-Molodetz} R. Penrose, ``Singularities and Time-Asymmetry''. In: Hawking, S.W. and Israel, W., Eds., General Relativity: An Einstein Centenary Survey, Cambridge University Press, Cambridge (1979) 581-638.

\bibitem{SH-Molodetz} S. W. Hawking,  ``A Brief History of Time: From the Big Bang to Black Holes''. New York: Bantam, 1988.

\bibitem{Hoyl-Narlikar-Molodetz} F. Hoyle and J. V. Narlikar, ``Time Symmetric Electrodynamics and the Arrow of Time in Cosmology'', doi: 10.1098/rspa.1964.0002
Proc. R. Soc. Lond. A (1964) 277, 1-23.

\bibitem{Wheeler-Feynman-Molodetz-1}   J.A. Wheeler  and R. Feynman,  ``Absorber Theory and the Radiation Arrow of Time'', Review of Modern Physics, 17 (1945) 157--181.

%\bibitem{Wheeler-Feynman-Molodetz-2}   J.A. Wheeler  and R. Feynman,  ''Classical Electrodynamics in Terms of Direct Interparticle Action'', Review of Modern Physics, 21 (1949) 425.

\bibitem{Hogarth-Molodetz}  J. E.Hogarth, ``Cosmological considerations of the absorber theory of radiation'', Proc. R. Soc. Lond. A267 (1962)  365--383.


\bibitem{Nobel}   Richard P. Feynman,  ``The Development of the Space-Time View of Quantum Electrodynamics (a Nobel lecture)'', December 11, 1965, https://www.nobelprize.org/prizes/physics/1965/feynman/lecture/

\bibitem{VilGar-Molodetz}     J. Garriga, A. Vilenkin, ``Testable anthropic predictions for dark energy'', Phys. Rev. D 67, 043503 (2003).


\bibitem{Carter} B. Carter, ``The anthropic principle and its implications for biological evolution'',  Phil. Trans. R. Soc. Lond. A310, 347 (1983).

\bibitem{Barrow_Tipler} J. D. Barrow and F. J. Tipler, ``The Anthropic Cosmological Principle'', Oxford University Press, New York (1986).

\bibitem{Weinberg} S. Weinberg, ``Anthropic Bound on the Cosmological Constant'', Phys. Rev. Lett. 59, 2607 (1987).

\bibitem{Linde-1} A. D. Linde, ``Inflation and quantum Cosmology'', {\em PRINT-86-0888-CAL-TECH} (1986). Published in ``300 Years of Gravitation'', Eds. S. W. Hawking and W. Israel, Cambridge University Press, Cambridge (1987)

\bibitem{Susskind-Molodetz}  Leonard Susskind, ``Three Impossible Theories'', arXiv:2107.11688  [hep-th].

\bibitem{ZelNov-Molodetz}  Ya. B. Zel'dovich, I. D. Novikov,  ``Relativistic Astrophysics, 2: The Structure and Evolution of the Universe'', Publisher: University Of Chicago Press, 1983.

\bibitem{Gold-Molodetz} T. Gold, ``The Arrow of Time'', American Journal of Physics (1962) 30, p. 403--410.

\bibitem{Hawking-Molodetz} S. W. Hawking,  ``Arrow of Time in Cosmology'', Physical Review D, (1985) 32, p. 2489.

\bibitem{Perlmutter} S. Perlmutter {\it et al.}, ``Discovery of a supernova explosion at half the age of the Universe'', Nature 391, 51--54 (1998).

\bibitem{Riess} A.G. Riess {\it et al.}, ``Observational Evidence from Supernovae for an Accelerating Universe and a Cosmological Constant'', Astron. J. 116, 1009--1038 (1998).

\bibitem{Landau1} L. D. Landau, E. M. Lifshitz, ``The Course of Theoretical Physics, vol. 1: Mechanics'' (3rd ed.), Butterworth-Heinemann (1976).

\bibitem{Landau2} L. D. Landau, E. M. Lifshitz, ``The Course of Theoretical Physics, vol. 2: The Classical Theory of Fields'' (4th ed.), Butterworth-Heinemann (1975)

\bibitem{Bogoliubov} N. N. Bogoliubov, D. V. Shirkov, ``Introduction to the Theory of Quantized Fields'' (3rd ed.), John Wiley and Sons Inc (1980)

\bibitem{Bjorken} J. D. Bjorken, S. Drell, ``Relativistic Quantum Mechanics'', McGraw-Hill (1964)

\bibitem{Ahlfors} Lars Ahlfors, ``Complex Analysis. An Introduction to the Theory of Analytic Functions of One Complex Variable'' (3rd ed.),  McGraw-Hill Book Co., New York (1978)

\bibitem{Harrison} Edward Robert Harrison, ``Darkness at Night: A Riddle of the Universe'', Harvard University Press (1987).

\bibitem{Albrecht-Molodetz}  Andreas Albrecht, Daniel Phillips,   ``Origin of probabilities and their application to the multiverse'',  Phys. Rev. D 90, 123514 (2014).

\bibitem{CKR} Y. Chen, S. Kumar, and B. Ratra, ``Determining The Hubble Constant From Hubble Parameter Measurements'', The Astrophysical Journal 835, 86 (2017)

\bibitem{LandauStat} L. D. Landau, E. M. Lifshitz, ``The Course of Theoretical Physics, vol. 5: Statistical Physics'' (3rd ed.), Butterworth-Heinemann (1980)

\bibitem{Lind-Molodetz} G. Lindblad, ``On the generators of quantum dynamical semigroups''. Commun. Math. Phys. 48, 119--130 (1976). https://doi.org/10.1007/BF01608499

\bibitem{Kal-Molodetz} A. O. Caldeira and A. J. Leggett, ``Influence of damping on quantum interference: An exactly soluble model'', Phys. Rev. A 31  1059 (1985)

\bibitem{Men-Molodetz} M. B. Mensky. ``Continuous Quantum Measurements and Path Integrals''. IOP Publishing, Bristol and Philadelphia, 1993.

\bibitem{Fey-Molodetz}R. P. Feynman, ``Space-Time Approach to Non-Relativistic Quantum Mechanics'',  Rev. Mod. Phys. 20, 367 (1948),

\bibitem{Price-Molodetz} Huw Price,   ``Recent Work on the Arrow of Radiation'',  [Preprint] (2005) http://philsci-archive.pitt.edu/2216/

\bibitem{BarrowSuperMolodetz} J. D. Barrow, A. A. H. Graham, ``New Singularities in Unexpected Places'', Int. J. Mod. Phys. D 24, 1544012 (2015).

\bibitem{Alla-Lera-Tema-Molodetz} A.A. Yurova, A.V. Yurov and V.A. Yurov,  ``What Can the Anthropic Principle Tell Us about the Future of the Dark Energy Universe'', Gravit. Cosmol. 25, 342--348 (2019).

\end{thebibliography}
\end{document}